\documentclass[fleqn,10pt]{wlscirep}

\usepackage{subfigure}
\usepackage[none]{hyphenat}

\newcommand{\cP}{{\mathcal{P}}}
\newcommand{\cN}{{\mathcal{N}}}
\newcommand{\bC}{\mathbf{C}}
\newcommand{\bu}{\mathbf{u}}

\title{Emergence of world-stock-market network}

\author[1]{M. Saeedian}
\author[2]{T. Jamali}
\author[1]{M. Z. Kamali}
\author[3]{H. Bayani}
\author[4,5]{T. Yasseri}
\author[1,6,7]{G.R. Jafari}
\affil[1]{Department of Physics, Shahid Beheshti University, G.C., Evin, Tehran 19839, Iran}
\affil[2]{School of Physics, Institute for Research in Fundamental Sciences (IPM), P.O. Box 19395-5531, Tehran, Iran}
\affil[3]{Faculty of physics, Tehran Science and Research Branch, Islamic Azad University, Tehran, Iran}
\affil[4]{Oxford Internet Institute, University of Oxford, 1 St Giles, OX13JS Oxford, UK}
\affil[5]{Alan Turing Institute, London, UK}
\affil[6]{The Institute for Brain and Cognitive Science (IBCS), Shahid Beheshti University, G.C., Evin, Tehran 19839, Iran}
\affil[7]{Center for Network Science, Central European University, H-1051, Budapest, Hungary}

\begin{abstract}
In the age of globalization, it is natural that the stock market of each country is not independent form the other markets. In this case, collective behavior could be emerged form their dependency together. This article studies the collective behavior of a set of forty influential markets in the world economy with the aim of exploring a global financial structure that could be called \emph{world-stock-market network}. Towards this end, we analyze the cross-correlation matrix of the indices of these forty markets using Random Matrix Theory (RMT). We find the degree of collective behavior among the markets and the share of each market in their structural formation. This finding together with the results obtained from the same calculation on four stock markets reinforce the idea of a world financial market. Finally, we draw the dendrogram of the cross-correlation matrix to make communities in this abstract global market visible. The dendrogram, drawn by at least thirty percent of correlation, shows that the world financial market comprises three communities each of which includes stock markets with geographical proximity.
\end{abstract}
\begin{document}

\flushbottom
\maketitle
%
%
\thispagestyle{empty}


\section*{Introduction}

There are different scales at which one can look at the economic world: global scale, country scale, etc. What is observed at the country scale is that the correlation among economic institutions of a country causes the emergence of that country's economy. This statement indicates the common feature of all financial structure:  some correlated financial units, at a scale, construct a financial structure at a larger scale. There are enough evidences such as the influence of a country recession into another countries that demonstrate correlation among the economy of different countries. Thus, considering each country as a financial unit, we expect to have a financial structure at the global scale whose constituents are stock markets of different countries. We call such an abstract structure ``world stock market''. Here, two questions arise: (i) how the existence of this global market can be ascertained, and (ii) what are its communities?

The first question that can be addressed, regarding to the main feature of every stock markets, is the emergent of collective behavior. Thus, if there exists a world stock market, one then should be able to show the collective behavior of that market's constituents. In the econophysics literature, the common approach for studying collective behavior is to analyze the cross-correlation matrix $\bC$ of stock returns using random matrix theory~\cite{m,b,m2,guh}. Since RMT describes a fully random system, any deviation from it contains information about the collective behavior among the market's constituents, see e.g.,~\cite{lcbp,pgra,pg,pb,Raj,Podobnik,namjmp,CSE}. Here, we propose a method based on RMT for measuring the collective behavior. In order to make randome matrices, we shuffle the non-diagonal elements of $\bC$. This procedure results in erasing the existing pattern of correlation among market's constituents, and hence removes collective behavior. Therefore, we expect to obtain valuable information about the collective behavior in a market by comparing statistical characteristics of $\bC$ with those of the shuffled $\bC$. Among all characteristics, we use participation ratio - a tool for estimating the number of significant participants in an eigenvector of a matrix~\cite{pb,Mobarhan} - and develop two new quantities called \emph{relative participation ratio} (RPR) and \emph{node participation ratio} (NPR) which will be described in the method section. The first quantity measures the degree of collective behavior in a market and the second one determines the share of each market's constituent in the measured collective behavior. RPR can be used for ranking different markets based on the degree of their collective behavior. NPR determines how much a market's component behaves independently of the collective behavior in the whole market. It can be used for ranking elements of a market according to their independence level. We apply the proposed method to the indices of forty influential markets in the world economy, from January 2000 to October 2015, in looking for a global financial structure. The results demonstrate the existence of such structure. In order to support our finding we show similarity between the world stock market and four of its markets including two developed and two emerging markets. One of the common characteristic of both the world stock market and these four markets is the presence of some constituents evolving almost independently of the other ones. 

In order to address the second question and to get a better perspective of the correlation effect in the world stock market, we use dendrogram analysis. The results show three main communities along with some isolated stock markets which are less affected by a crisis and more affected by a booming.


\begin{table*}[t!]
\centering
\footnotesize
\begin{tabular}{ |p{5cm}|p{2cm} || p{5cm}|p{2cm}| }
\multicolumn{4}{c}{\normalsize{Markets List}} \\
\hline
\textbf{Market Name}  & \textbf{Country} &  \textbf{Market Name}  & \textbf{Country} \\
\hline
Buenos Aires Stock Exchange (MERVAL) & Argentina &  Mexican Stock Exchange (MXX) & Mexico \\ \hline
Australia Stock Exchange (S\&P/ASX 200 VIX) & Australia & Amsterdam Stock Exchange (ASE) & Netherland \\ \hline
Austrian Traded Index	(ATX) &	Austrian & Oslo Stock Exchange (OSE) & Norway \\ \hline
 Brussels Stock Exchange (BEL 20) & Belgium & Karachi Stock Exchange Limited (KSE) & Pakistan\\ \hline
Bolsa de Valores, Mercadorias \& Futuros	(BM\&F Bovespa) & Brazil & Bolsa de Valores de Lima (BVL) & Peru \\ \hline
Toronto Stock Exchange (S\&P/TSX) & Canada & Philippine Stock Exchange (PSE) & Philippine \\ \hline
Shanghai Stock Exchange (SSE)$^*$ & China & Qatar Stock Exchange (QSE) & Qatar \\ \hline
Copenhagen Stock Exchange (OMX Copenhagen 20) & Denmark  & Moscow Interbank Currency Exchange (MICEX) & Russia \\ \hline
Financial Times Stock Exchange (FTSE)$^*$ & England  & Tadawul All-Share Index (TASI) & Saudi Arabia \\ \hline
French stock market (CAC 40) & France &  Johannesburg Stock Exchange (JSE Limited) & South Africa \\ \hline
Deutscher Aktienindex (DAX) & Germany & \'Indice Burs\'atil Espa\~nol (IBEX) & Spain \\ \hline
Hang Seng Index (HIS) & Hong Kong & Colombo Stock Exchange (CSE) & Sri Lanka \\ \hline
Bombay Stock Exchange (BSE) & India & Stockholm Stock Exchange (OMX Stockholm 30) & Sweden \\ \hline
Jakarta Stock Exchange (JSX) & Indonesian  & Swiss Market Index (SMI) & Switzerland \\ \hline
Tehran Stock Exchange (TSE)$^*$ & Iran & Taiwan Stock Exchange Corporation (TWSE) & Taiwan  \\ \hline
Irish Stock Exchange (ISE) & Ireland & Borsa Istanbul (BIST) & Turkey \\ \hline
Borsa Italiana (SpA) & Italy & Dow Jones Industrial Average (DJIA) & USA \\ \hline
Tokyo Stock Exchange (Nikkei 225) & Japan & Standard \& Poor's 500 (S\&P 500)$^*$ & USA \\ \hline
Korea Composite Stock Price Index (KOSPI) & Korea & Nasdaq Stock Market (NASDAQ) & USA \\ \hline
FTSE Bursa Malaysia KLCI	(FBM KLCI)	& Malaysia & Zurich Stock Exchange (ESTX50) & Zurich \\ \hline
\end{tabular}
\label{tab: markets list}
\caption{\label{tab: markets list} Names of forty stock markets and their corresponding countries, listed according to the alphabetical order of countries name.  Among all the stock markets throughout the globe, we choose these markets with regards to their GDP and geographical considerations. The four markets indicated by asterisk sign are studied twice: once as components of a global market and once as independent markets.}
\end{table*}

\section*{Result and Discussion}

Here, the proposed method of this study is first applied to the forty stock market indices to trace a global economic structure. These markets together with their corresponding countries are listed in Tab.~\ref{tab: markets list}. The markets are chosen due to Gross Domestic Product (GDP) and geographical considerations. We then apply our method to four stock markets, indicated by asterisk sign in Tab.~\ref{tab: markets list}, to observe similar structure but at the lower scale.

\subsection*{Forty markets as a world stock market}

The common approach in studying collective behavior of a market is based on RMT results and its deviation from market results. Recently, two of the authors introduced another criterion based on fractional Gaussian noises~\cite{jj}. The eigenvalue distribution of markets differs from RMT's distribution; there are some eigenvalues out of the RMT bulk region. These deviating eigenvalues contains useful information about the collective behavior. To be more precise, it was shown that large eigenvalues show the markets' trend and the largest eigenvalue indicates the largest collective mode in markets, see e.g., \cite{pb,Raj,namjmp,nampa}. In the following, we study collective behavior in an abstract market named ``world stock market'' by taking a different path using the proposed method of this paper.

\begin{figure}[t!]
\centering
    \subfigure{
        \centering
        \includegraphics[width=0.4\textwidth]{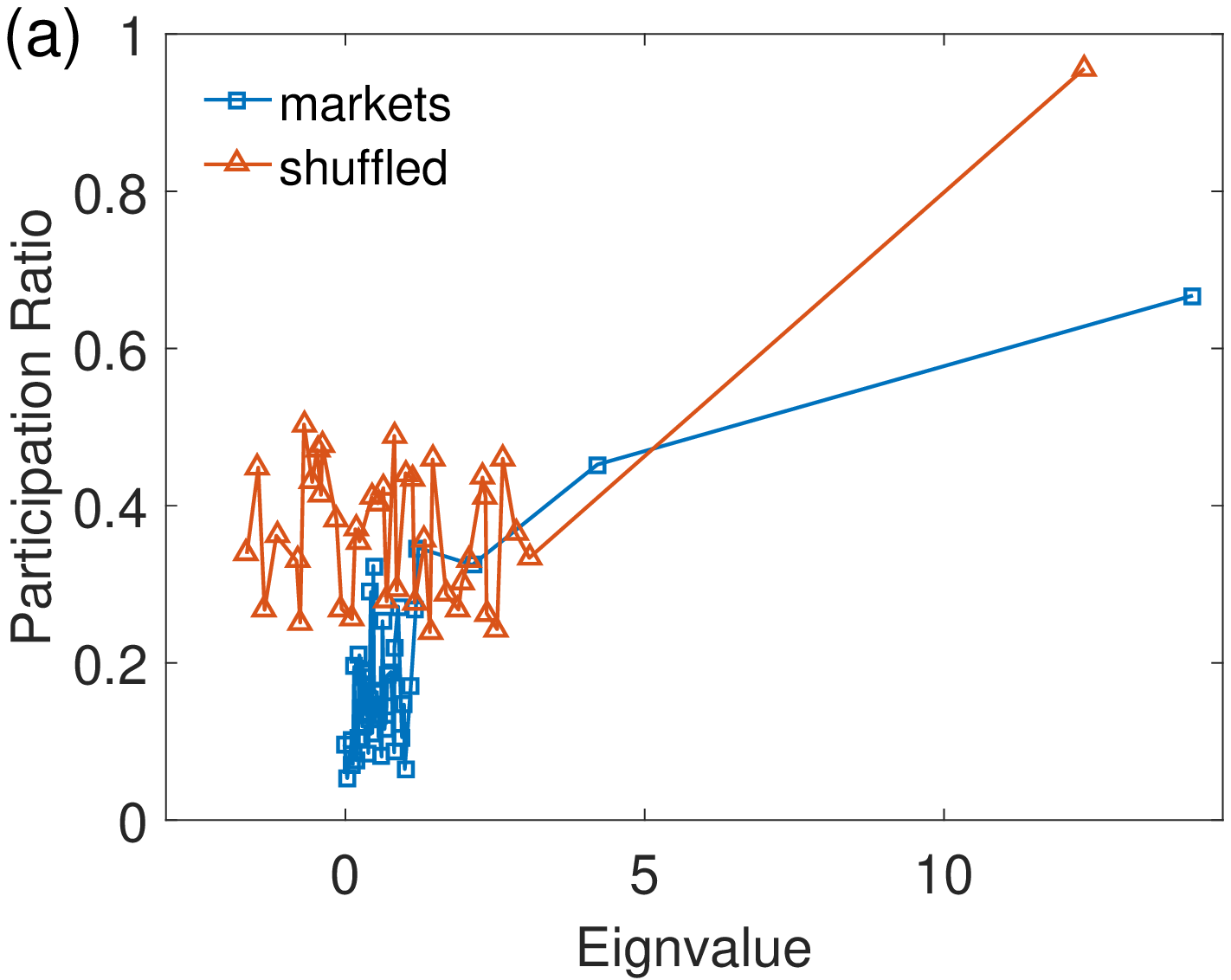}\label{f5}
    }
    ~
    \subfigure{
        \centering
        \includegraphics[width=0.4\textwidth]{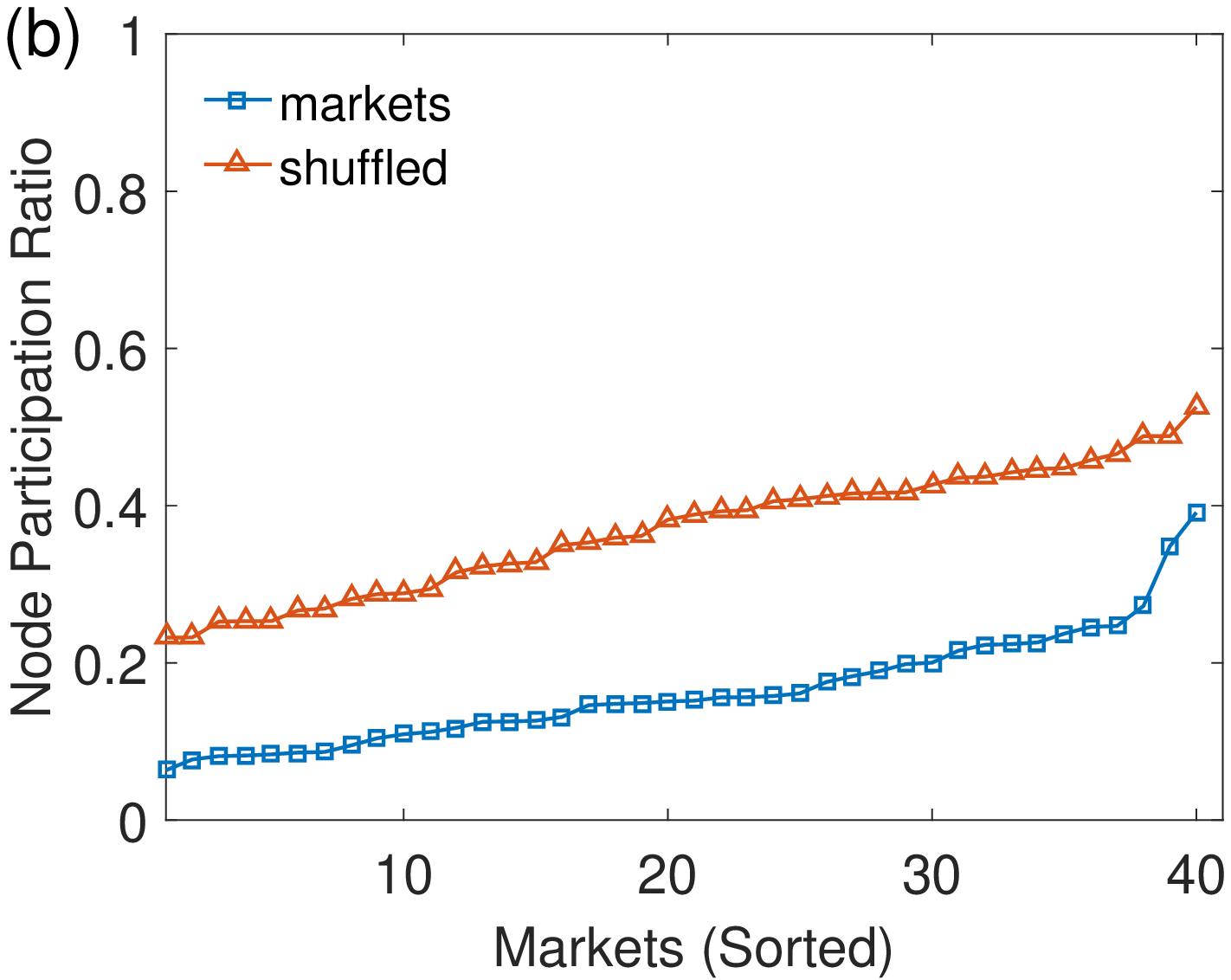}\label{f6}
    }
\caption{(a) Participation ratio and (b) node participation ratio of the world stock market consisting of the forty markets of Tab.~\ref{tab: markets list} along with corresponding shuffled ratios.}
\end{figure}

Assuming that there exists a world stock market whose constituents are the forty markets, one can then construct the corresponding cross-correlation matrix $\bC$ and its shuffled counterpart $\bC_{sh}$. For this purpose, we use the indices data of these forty markets in the period January 2000 to October 2015~\cite{yah}. After diagonalizing $\bC$ and $\bC_{sh}$,  participation ratios $\cP_k$ are then obtained using Eq.~(\ref{pr}). Figure~\ref{f5} shows the PRs of the world stock market and its shuffled version. As seen from this figure, the PR of the shuffled are greater than the market in average yielding to the relative participation ratio $\delta \approx 0.5$. This number represents the degree of collective behavior among the forty markets. In the next subsection, we will do the same calculation for four stock markets.


After showing that there is a collective behavior among these 40 markets, we find the share of each market in that behavior using the NPR parameter, Eq.~(\ref{6}). Figure~\ref{f6} depicts the NPR for the world stock market in which 40 markets are sorted ascendingly according to their NPR values. The markets on the left side are more independent from the trend of the world stock market while the ones on the right side are more dependent on the market's collective behavior. This figure also shows the effect of shuffling on NPR. The notable result here is that the markets with low NPRs, located in the left side of Fig.~\ref{f6}, reduce the risk of a world portfolio because they have higher independence level than other markets and hence at the time of crashes they will be less affected by the world trend.

\begin{figure}[t!]

\centering
    \subfigure[]{
        \centering
        \includegraphics[width=0.37\linewidth]{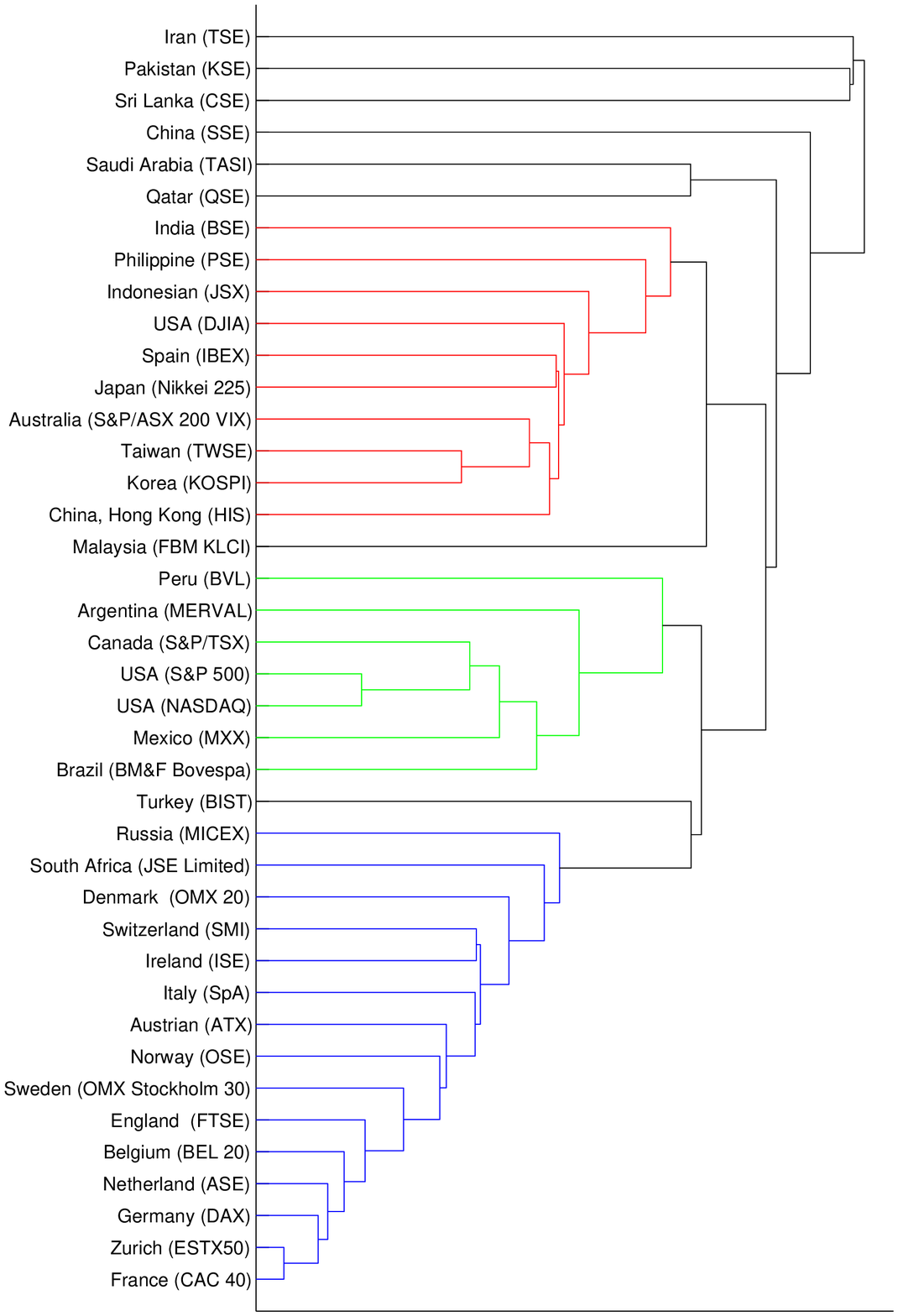}\label{dendrogram}
    }
    ~
    \subfigure[]{
        \centering
        \includegraphics[width=0.58\linewidth]{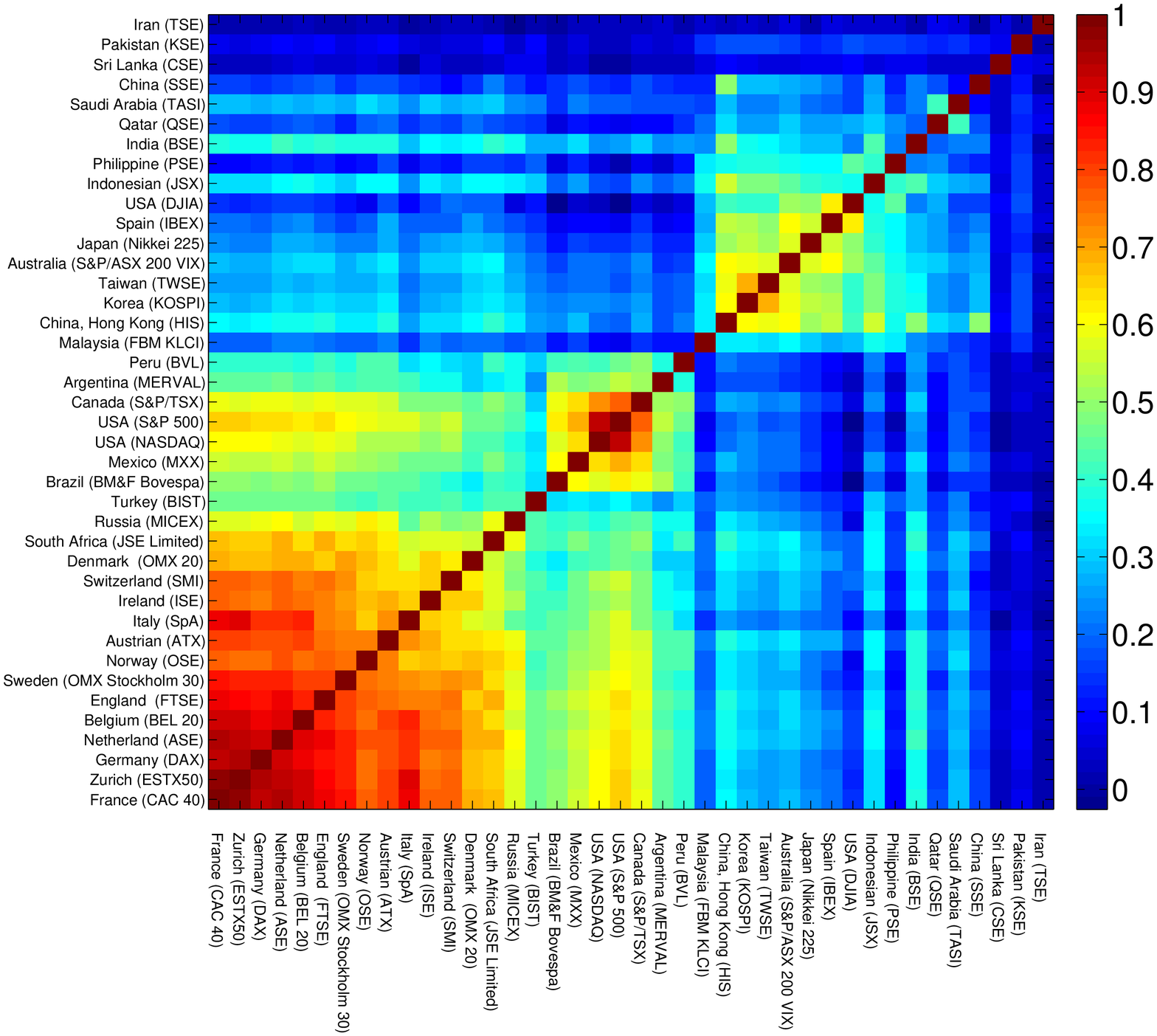}\label{correlation matrix dendrogram}
    }
\caption{(a) Dendrogram of the forty markets in Tab.~\ref{tab: markets list} based on their cross-correlations. The colored groups contain members with at least thirty percent of correlation. The top six markets do not belong to any group. (b) heatmap of cross-correlation between forty markets. Dark red (blue) regions shows strong (weak) correlations.
}
\end{figure}

Figure~\ref{dendrogram} is the dendrogram of the cross-correlation matrix of the world stock market after financial crash 2008. This dendrogram shows three communities in the world market of 40 indices, colored by red, green and blue, according to their correlation distances. The clustering is highlighted by at least thirty percent of correlation between the markets. Looking at the component of these communities illustrates the effect of geographical relations between them. The red, blue and green communities are mostly consist of the markets located in East Asia, Europe and the continent of America, respectively. The black color markets are those with less than thirty percent of correlation. These markets, which are the ones with less NPRs, are Asian countries, namely China, Iran, Pakistan, Qatar, Saudi Arabia, and Sri Lanka. Figure~\ref{correlation matrix dendrogram} is the corss-correlation matrix of the world stock market where its rows and columns are rearranged according to the dendrogram pattern of Fig.~\ref{dendrogram}. The color of each square cell represents the value of cross-correlation between the two markets. Three communities, around the secondary diagonal of the matrix, can be clearly observed.

\subsection*{Four markets}

\begin{figure}[t!]
\centering
    \subfigure{
        \centering
        \includegraphics[width=0.4\textwidth]{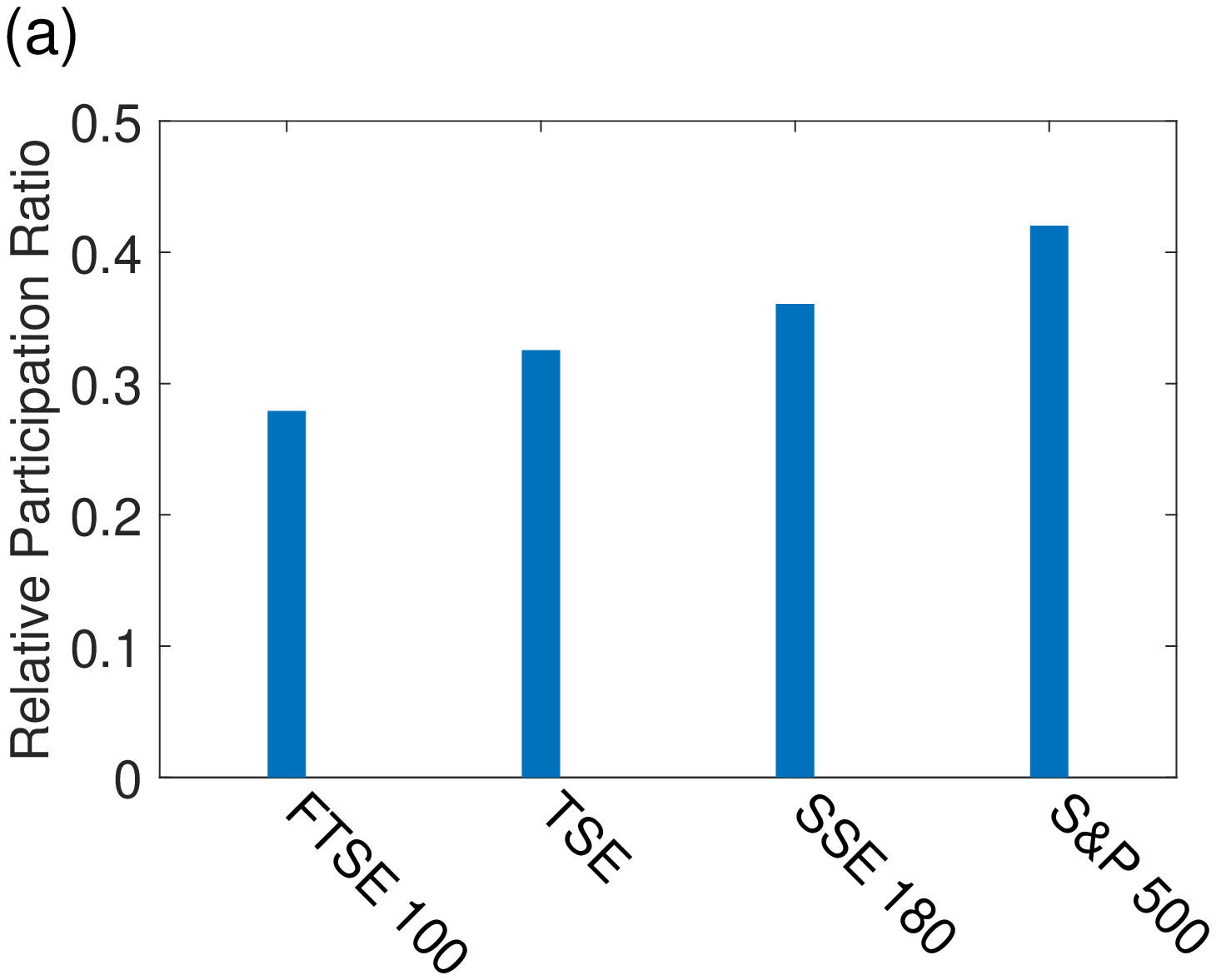}
        \label{f2}
    }
    ~
    \subfigure{
        \centering
        \includegraphics[width=0.4\textwidth]{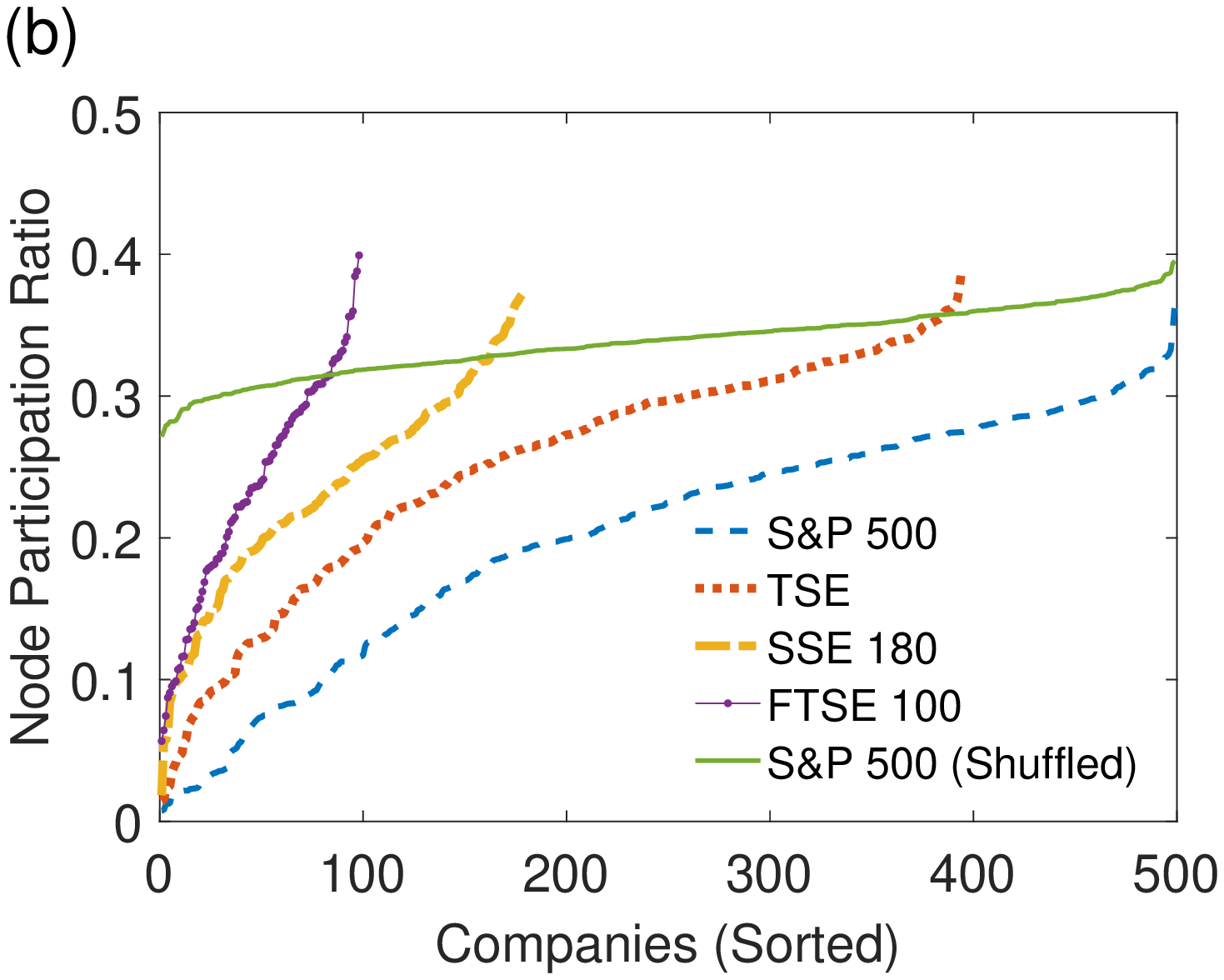}
        \label{f3}
    }
\caption{(a) Relative participation ratio of the four markets. The value of RPR indicates the level of collective behavior in a market. This figure illustrates that the collective behavior does not depend on the type of market, i.e., emerging or developed. (b) Node participation ratio of the four markets. The horizontal axis is the companies of markets sorted according to their NPR values. In order to show the effect of shuffling on NPR and to have a better comparison with the NPR of the world stock market, Fig.~\ref{f6}, we also draw the NPR of shuffled S\&P 500. 
}
\end{figure}

Here, we apply the proposed method to the indices of the four stock markets includes the Standard \& Poor's 500 (USA) and the Financial Times Stock Exchange 100 (United Kingdom) as developed markets and the Shanghai Stock Exchange 180 (China) and the Tehran Stock Exchange (Iran) as emerging markets. The data of the markets are in the same period as 40 markets. Figure~\ref{f2} shows the relative participation ratio, $\delta$, for these markets. Note that in order to have a correct comparison between different markets, before using Eq.~\ref{5}, the PRs of these markets are normalized by the markets' size. Since $\delta$ represents the degree of collective behavior in a market, Fig.~\ref{f2} shows that the companies of S\&P 500 have the highest degree of collective behavior among the four markets. This can be interpreted in this way that a strong collective atmosphere exists in S\&P 500. Fig~\ref{f3} shows the normalized NPR of each market in a sorted fashion like Fig.~\ref{f6}. This figure gives the share of each company in a collective behavior of market. The other notable point is that the degree of collective behavior, $\delta$, does not depend on the type of market for instance although SSE 180 and TSE are emerging markets but they have a greater $\delta$ than that of the developed market FTSE 100. 

\begin{figure}[t!]
\centering
    \subfigure{
        \centering
        \includegraphics[width=0.22\textwidth]{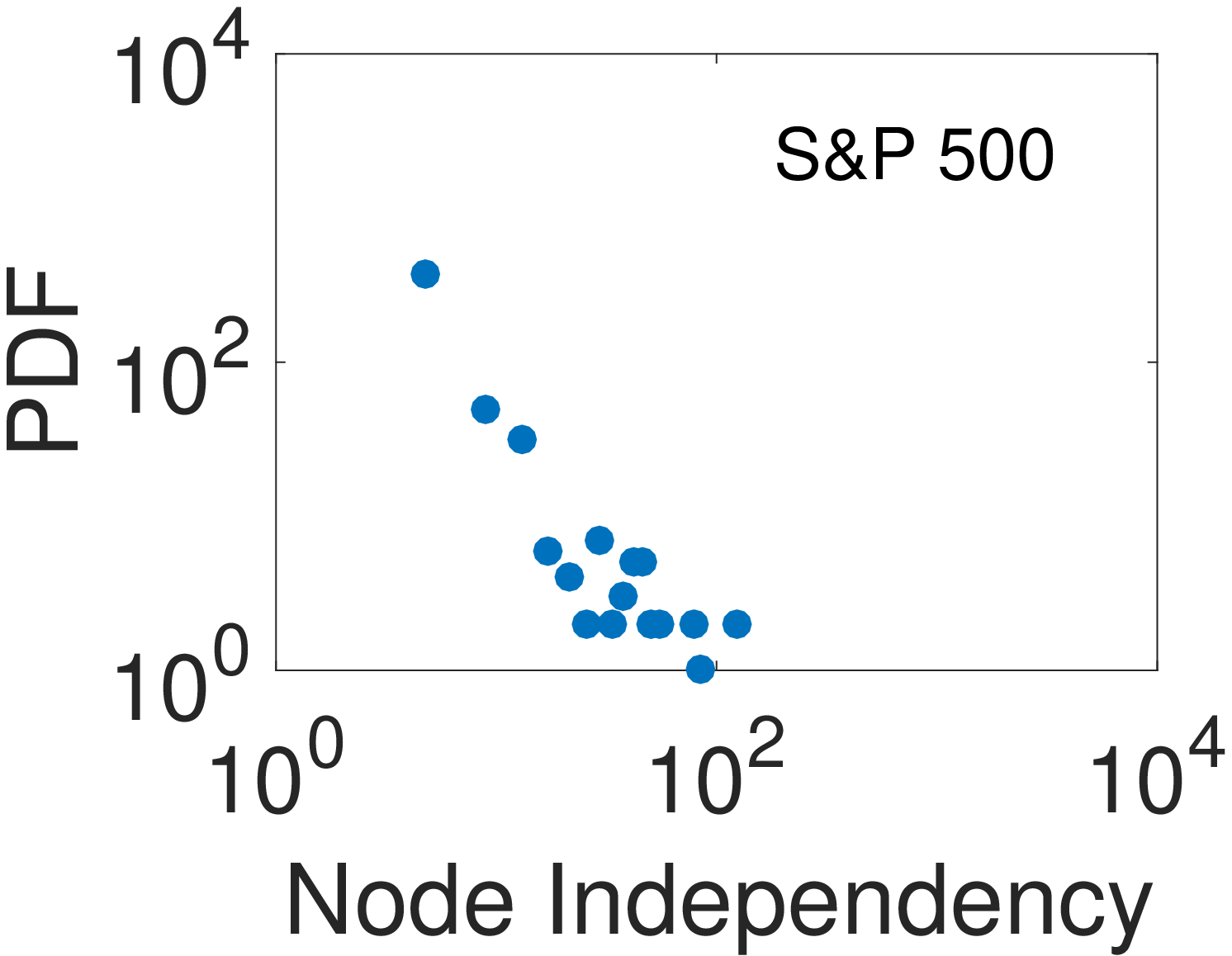}
        \label{fig:subfigure1}
    }
    ~
    \subfigure{
        \centering
        \includegraphics[width=0.22\textwidth]{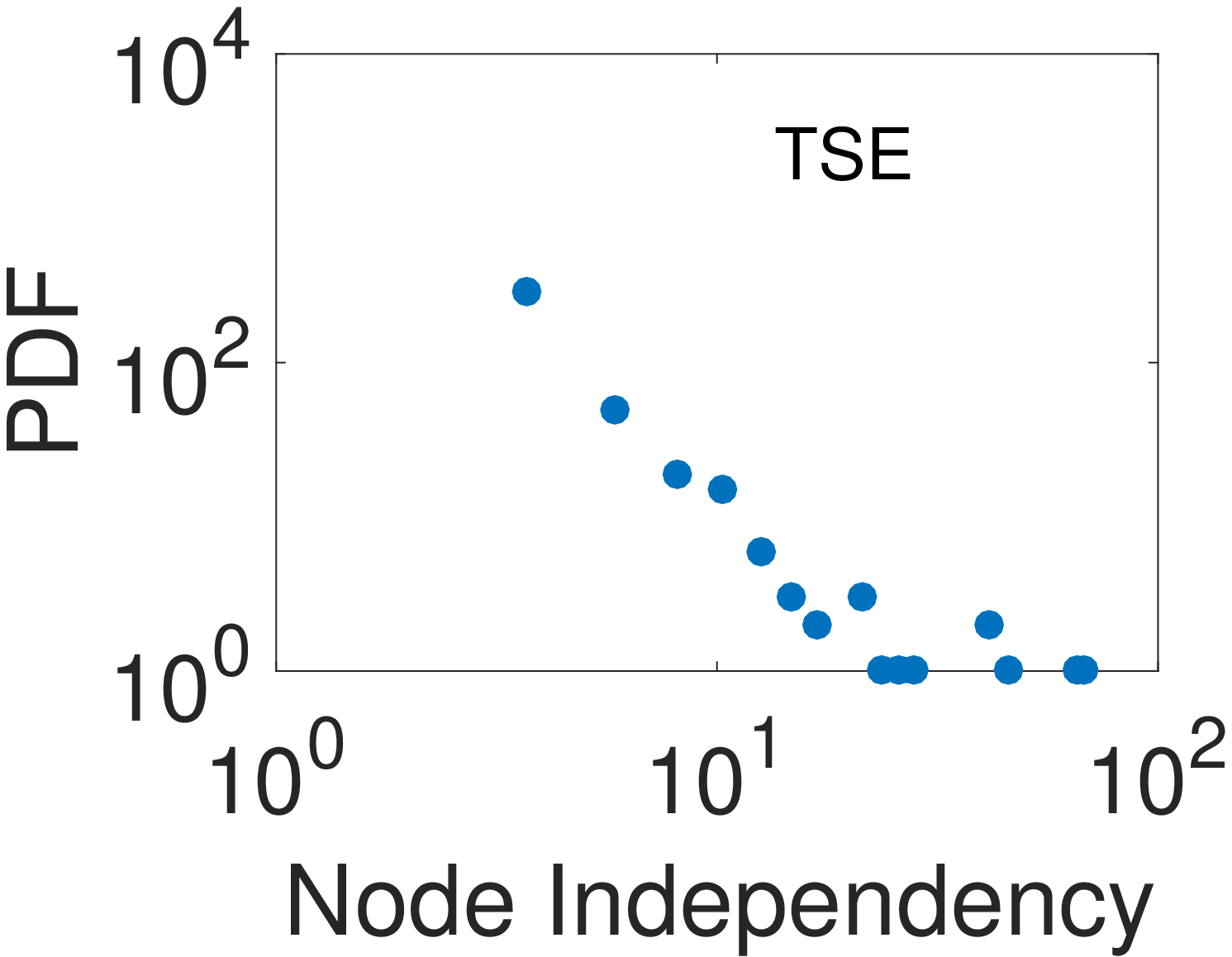}
        \label{fig:subfigure2}
    }

    \subfigure{
        \centering
        \includegraphics[width=0.22\textwidth]{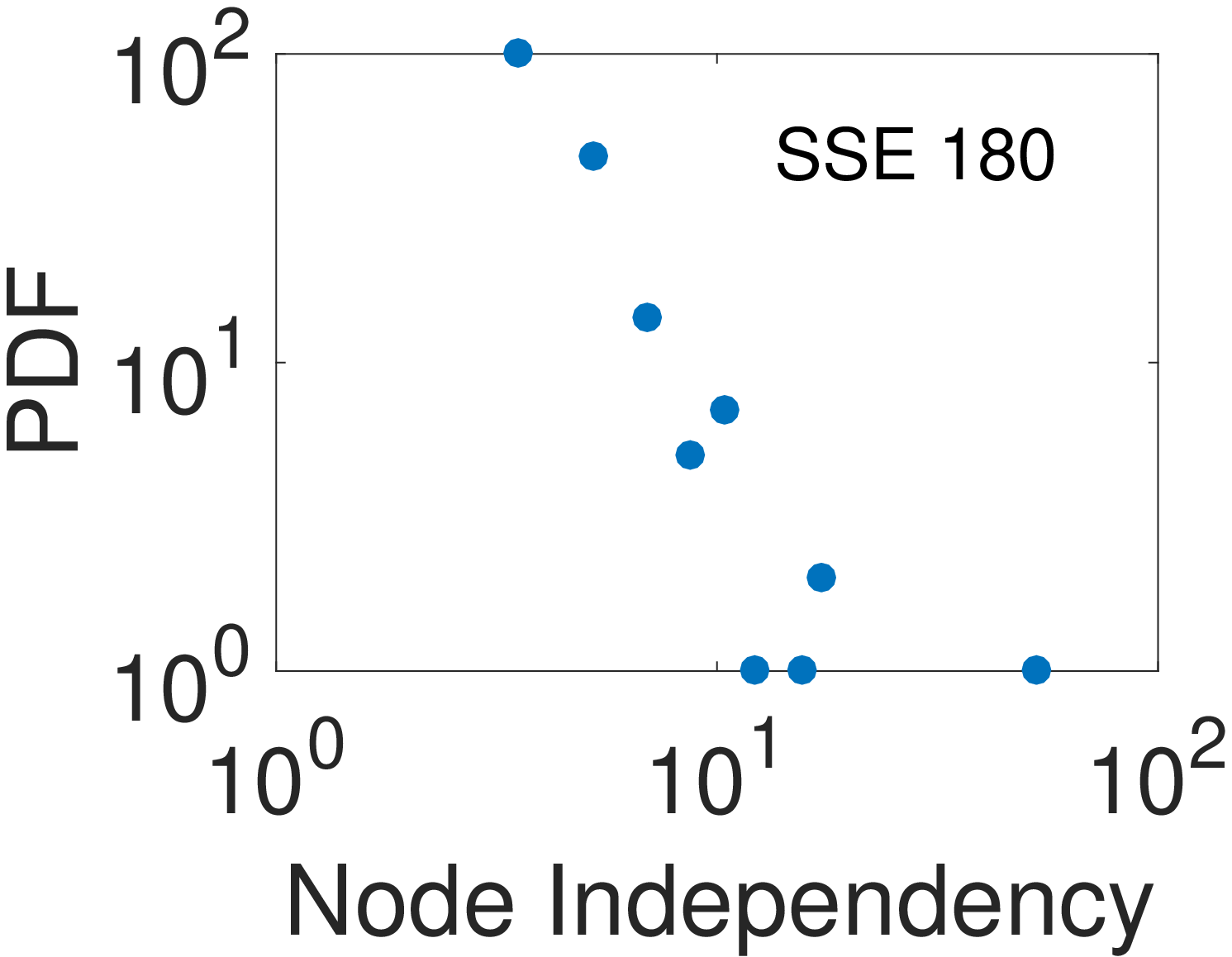}
        \label{fig:subfigure3}
    }
    ~
    \subfigure{
        \centering
        \includegraphics[width=0.22\textwidth]{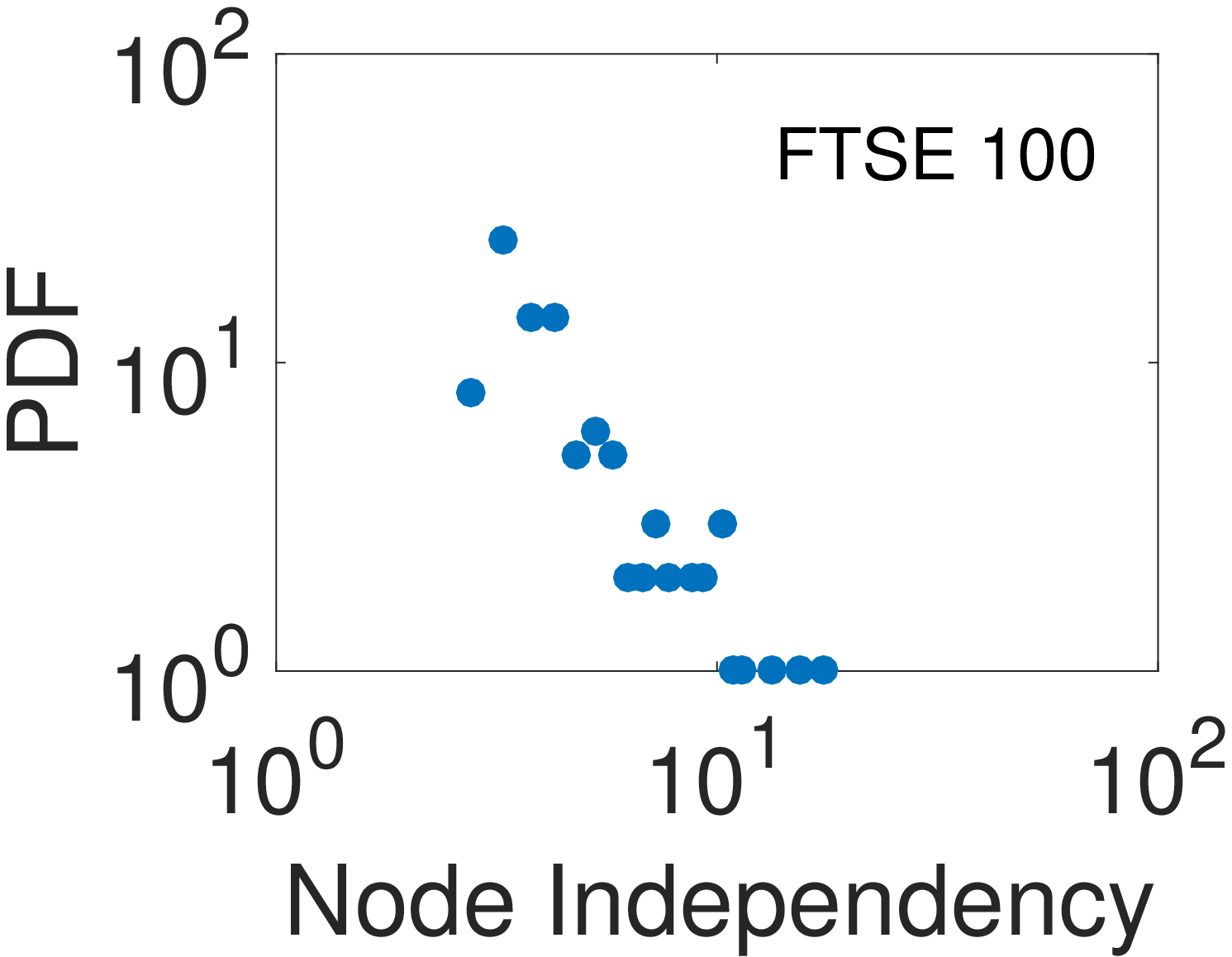}
        \label{fig:subfigure4}
    }
\caption{The log-log plots of independency distribution, which is the inverse of node participation ratio, for the four stock markets. This figure shows fat-tailed behavior for the contribution of companies in the collective behavior. The tails of the PDF correspond to the few number of companies with high degree of independency from the collective behavior. The most of companies participate considerably in the collective behavior.
}
\end{figure}

The green solid line in Fig~\ref{f3}, represents the effect of shuffling on S\&P500 NPR which is exactly similar to what is observed in Fig.~\ref{f6}. In order to identify the contribution of companies in the collective behavior more clearly, we also compute the probability density function (PDF) of node independency, which is the inverse of NPR. Figures~\ref{fig:subfigure1}-~\ref{fig:subfigure4} are the PDFs of node independency and interestingly illustrate fat-tail behavior. This means that there are very few companies in each of these markets that work almost independently and have small impact on the collective behavior, while most companies contribute in the collective behavior remarkably.

%


\section*{Summary and Conclusion}

Historically, stock markets were emerged from the central sovereign states and territories. However, in the globalization age, stock markets have been severely affected by communications, so that the future and the existence of all countries tie together. In this work, we have studied the network of forty influential markets from different countries to address this question that whether the globalization results in the emergence of a world stock market. Due to this fact that every financial system consists of many units with collective behavior, we expect to observe such behavior for the world stock market whose units are these forty markets. In order to meet this expectation, a method have been introduced for measuring collective behavior in a market. This method is based on the concept of participation ratio. We have shown that the forty markets possess collective behavior and their shares in this collective behavior are not the same. The community of the forty markets have also been extracted using the dendrogram technique; the result shows three main communities plus some isolated markets belong to some Asian countries. These markets have the lowest shares in the global collective behavior or in other words have the highest level of independency from the global trend. The three communities, on the other hand, have more participation in the global collective behavior. Moreover, each of these communities includes markets belonging to countries with geographical proximity. Eventually, the results of this study illustrates the collective behavior among forty markets and therefore proves the existence of a world stock market.

\section*{Method}

Here, we present a method based on random matrix theory. Historically, this theory traces back to the work of Wigner in nuclear physics where the precise nature of the interactions between the components of atomic nuclei are not known~\cite{epa,epb,epc,fj1,fj2,m1}. From the viewpoint of having unknown underlying interactions, financial systems are very similar to atomic nuclei. Laloux et. al. \cite{lcbp} demonstrated that RMT could be a suitable candidate for studying financial correlation matrices; then, Plerou et. al.~\cite{pgra} extract statistical properties of cross correlations in financial data using RMT. In order to construct the cross-correlation matrix $\bC$, the price return of the $i$th stock is first calculated as
\begin{equation} \label{1}
{R_i}(t) = \ln {P_i}(t+\Delta t) - \ln {P_i}(t),
\end{equation}
where $i=1,\dots,N$, $\Delta t$ is the time scale, and $P_i(t)$ indicates the price of the $i$th stock. Since the returns of stocks have different variances, it is suitable to work with the normalized price return $r_i(t)$, instead of $R_i(t)$, which is defined as
\begin{equation}
\label{2}
{r_i} = \frac{{R_i}(t) - \left\langle {R_i} \right\rangle_t }{\sigma _i},
 \end{equation}
where  ${\sigma _i} = \sqrt{\langle R_i^2 \rangle_t  - \langle R_i \rangle_t^2}$ is the standard deviation of the return $R_i(t)$, and $\left\langle \cdots \right\rangle_t $ indicates the time average over the period of study. The equal-time cross-correlation matrix $\textbf{C}$ is then constructed with the elements $C_{ij}$ given by
\begin{equation}
 \label{3}
  {C_{ij}} = \langle {{r_i}(t)\,{r_j}(t)} \rangle_t.
 \end{equation}
From Eqs. (\ref{2}) and (\ref{3}) it is readily seen that $\bC$ is a symmetric matrix with unit diagonal elements and off-diagonal elements in $[-1,1]$. In the following subsections, we present two quantities for the measurement of collective behavior among markets based on the cross-correlation matrix  $\bC$; but, before that we introduce another matrix named ``shuffled cross-correlation matrix'' which is the counterpart of $\bC$, and state its potential  application in this context.

The cross-correlation matrix $\bC$ can be diagonal, which means that there is no interaction or correlation between the markets, or off-diagonal, which, on the contrary, means that there is correlation between markets. The existence of correlation is a necessary condition, but not sufficient, for the emergence of collective behavior among markets. This statement can be justified in this way that we do not expect to observe a collective behavior in a stock market whose  constituents are correlated to each other in a completely random fashion. Thus, besides having correlation between markets, a sort of pattern or structure for that correlation is needed. In other words, collective behavior is emerged when there exists a structure for the market in addition to the correlation among the market's constituents. Now a question arises: how to make visible such a structure or its effect? To answer this question, we randomly shuffle the off-diagonal elements of $\bC$. The new matrix obtained in this way is called  \emph{shuffled} cross-correlation matrix and denoted by $\bC_{sh}$. Note that random shuffling the off-diagonal elements would vanish any specific pattern of correlation without annihilating correlations themselves. Briefly, two matrices can be assigned to each market: the cross-correlation matrix $\bC$ containing both the correlation values and structure and the shuffled cross-correlation matrix $\bC_{sh}$ containing only correlation values.

\subsection*{Relative participation ratio}

In order to quantify the degree of collective behavior in a market, we introduce a quantity based on the concept of participation ratio (PR) which is first defined by Bell and Dean \cite{db} in the context of atomic physics. Diagonalizing $\bC_{N\times N}$ gives us a set of eigenvectors $\{\bu_k\}$ and eigenvalues $\{\lambda_k\}$. Note that an eigenvalue represents a collective mode of market and its corresponding eigenvector contains the share of market's components in that collective mode. For the $k$th eigenvector, participation ratio is defined as follows:

 \begin{equation}
 \label{pr}
\cP_k \equiv \left(\sum\limits_{l = 1}^N [u_k(l)]^4 \right)^{ - 1}
 \end{equation}
where $u_k(l), l=1,\dots,N$  are the components of $\bu_k$. Participation ratio $\cP_k$ is bounded from below by unity for the case of $\bu_k$ with only one non-zero component and from above by $N$ for the case of $\bu_k$ with identical components $u_k(l) = N^{-1/2}$. This gives the natural meaning of the PR as a measure for the number of significant components in an eigenvector. Since PRs of a market depends on its size $N$, a correct comparison between the PRs of various markets of different sizes could be obtained when PRs become size independent. For this purpose, we normalize PRs, Eq.~(\ref{pr}), by the size of market so that the maximal bound of $\cP_k$ becomes unity.

According to the reason for the construction of $\bC_{sh}$, we now define a new parameter named \emph{relative participation ratio} (RPR) as follows
 \begin{equation} \label{5}
\delta = \frac{\langle \cP_{sh} \rangle - \langle \cP \rangle}{\langle \cP_{sh} \rangle},
 \end{equation}
where $\langle \cP_{sh} \rangle$ and $ \langle \cP \rangle$ represent the average of PRs over all eigenvectors of $\bC_{sh}$ and $\bC$, respectively. Since the parameter $\delta$ quantifies the deviation of the participation ratio of the cross-correlation matrix from its shuffled counterpart in an average sense, it gives us the degree of collective behavior pattern in a market. When there is a week collective behavior in the market, random shuffling should has small effect on $\bC$, i.e., $\langle \cP \rangle \approx \langle \cP _{sh} \rangle$, and hence $\delta$ is near zero. On the other hand, when a strong pattern of collective behavior presents, random shuffling has considerable effect on $\bC$, and consequently we have a large $\delta$.

\subsection*{Node participation ratio}
Quantifying the collective behavior in a market, this question may arise that how one can specify the contribution of each market's constituent in the measured collective behavior. To address this question, we introduce a new quantity, named \emph{node participation ratio} (NPR), as follows
 \begin{equation} \label{6}
\cN_l \equiv \left(\sum_{k = 1}^N [u_k(l)]^4\right)^{-1}.
 \end{equation}
Notice that the summation is taken over the index ``$k$'', i.e., over the $l$th row of eigenvectors. Since the eigenvector $\bu_k$ includes the share of market's components in the collective mode related to the eigenvalue $\lambda_k$, the NPR $\cN_l$ determine the share of the $l$th component in the total collective behavior. In the language of stock markets, this quantity can be also interpreted in this way that a company with lower NPR evolves more independently than a company with higher NPR. As a result $\cN_l^{-1} $ gives a measure of the independence of the $l$th company from other companies.

\bibliographystyle{unsrt}
\bibliography{Draft}

\section*{Acknowledgments}
GRJ and MS gratefully acknowledge support from Cognitive Science and Technologies Council grant No. 2694.

%

%


\end{document}